\documentclass[apl,twocolumn]{revtex4-1}

\usepackage{amssymb,amsmath}
\usepackage{microtype}
\usepackage{mathrsfs}
\usepackage{hyperref}
\usepackage{tikz}

\tikzset{cross/.style={path picture={
      \draw[black]
            (path picture bounding box.south east) --
            (path picture bounding box.north west)
            (path picture bounding box.south west) --
            (path picture bounding box.north east);}}}

\usetikzlibrary{intersections}

\newcommand{\pd}{\partial}

\begin{document}

\date{\today}

\title{A challenge to the $a$-theorem in six dimensions}

\author{Benjam\'in Grinstein}
\email{bgrinstein@ucsd.edu}
\affiliation{Department of Physics, University of
California, San Diego, La Jolla, CA 92093, USA}
\author{Andreas Stergiou}
\email{andreas.stergiou@yale.edu}
\affiliation{Department of Physics, Yale University, New Haven, CT 06520,
USA}
\author{David Stone}
\email{dcstone@ucsd.edu}
\affiliation{Department of Physics, University of
California, San Diego, La Jolla, CA 92093, USA}
\author{Ming Zhong}
\email{zhongm@nudt.edu.cn}
\affiliation{Department of Physics, University of
California, San Diego, La Jolla, CA 92093, USA}
\affiliation{Department of Physics, National University of Defense
Technology, Hunan 410073, China}

\begin{abstract}
  The possibility of a strong $a$-theorem in six dimensions is examined in
  multi-flavor $\phi^3$ theory. Contrary to the case in two and four
  dimensions, we find that in perturbation theory the relevant quantity
  $\tilde{a}$ increases monotonically along flows away from the trivial
  fixed point.  $\tilde{a}$ is a natural extension of the coefficient $a$
  of the Euler term in the trace anomaly, and it arises in any even
  spacetime dimension from an analysis based on Weyl consistency
  conditions.  We also obtain the anomalous dimensions and beta functions
  of multi-flavor $\phi^3$ theory to two loops.  Our results suggest that
  some new intuition about the $a$-theorem is in order.
\end{abstract}

\maketitle

\section{Introduction}
The counting of degrees of freedom in quantum field theories (QFTs) is of
paramount importance in understanding their structure and phases. In
particular, it is often of interest to understand how low-energy,
long-range ``IR'' degrees of freedom might be related to the underlying
microscopic ``UV'' degrees of freedom. For example, in quantum
chromodynamics we observe mesons, hadrons, {\em etc.}\ at low energies, but
believe them to consist of the quarks and gluons of the microscopic theory.

A rather good understanding of QFT degrees of freedom exists in two
dimensions. There, a quantity can be defined that undergoes a monotonically
decreasing renormalization group (RG) flow from a critical point in the UV
to a critical point in the IR. At the critical points the quantity is
stationary with respect to variations in scale, and becomes the central
charge $c$ of the Virasoro algebra of the corresponding conformal field
theory (CFT), which is also the coefficient of the topological term (the
Ricci scalar) in the two-dimensional trace anomaly. This is the result of
Zamolodchikov~\cite{Zamolodchikov:1986gt}.

In the four-dimensional case, which is of great interest to particle
physicists, results are not so definitive. Cardy
suggested~\cite{Cardy:1988cwa} that the four-dimensional analog of $c$ is
the coefficient of the (topological) Euler term in the four-dimensional
trace anomaly, $a$.  In fact, it was shown, using heat kernel methods for
field theories on curved backgrounds~\cite{Jack:1983sk} and Weyl
consistency conditions~\cite{Osborn:1991gm}, that a perturbative version of
Zamolodchikov's result holds~\cite{Jack:1990eb, Osborn:1991gm}.  More
recently, non-perturbative methods have made headway into a weaker version
of the $a$-theorem, where, instead of establishing a monotonic flow, a
relation between the value of $a$ at the critical points is argued
for~\cite{Komargodski:2011vj}, namely that $a_{\text{UV}}>a_{\text{IR}}$.

In this paper we investigate the possibility of an $a$-theorem in six
dimensions. The weak version of the $a$-theorem in $d=6$ was studied
in~\cite{Elvang:2012st} using the methods of~\cite{Komargodski:2011vj}, but
no definitive conclusion could be reached. The six-dimensional case is of
interest in clarifying the basic structure of QFT in general. Interesting
CFTs arise in $d=6$ by string-theoretic constructions and the low energy
dynamics of M5 branes.  The spectrum of operators in these theories can be
studied without any knowledge of what Lagrangian ``describes'' them, but
not much is known about RG flows to and from these theories. As far as
Lagrangian theories are concerned, the $\phi^3$ theory is of interest as
the unique classically scale-invariant theory in $d=6$. Its RG running can
be easily studied with well-known methods, and expectations stemming from
the intuition behind the $a$-theorem can be put to the test.

In this Letter we examine the possibility of an $a$-theorem in
six-dimensional multi-flavor $\phi^3$ theory. We find that the opposite
conclusion of two- and four-dimensional $a$-theorems may be drawn in six
dimensions, at least in perturbation theory. More specifically, we find
that the candidate for an $a$-theorem singled out by the Weyl consistency
conditions {\em increases} monotonically along the renormalization group
flow out of the trivial fixed point. To come to our conclusion, we use the
methods developed in \cite{Osborn:1991gm} (see also \cite{Jack:2013sha,
Baume:2014rla}). This involves constraining the form of the Weyl anomaly
utilizing the Abelian nature of the Weyl group; because the Weyl group is
related to a change of scale, this imposes constraints on the RG properties
of quantities in the anomaly and, in particular, produces a candidate for
an $a$-theorem. In section~\ref{sec:weylccs} we explain this method and in
section~\ref{sec:results} we show that a quantity that becomes $a$ at
critical points {\em increases} monotonically along the renormalization
group flow, at least in perturbation theory. We discuss the implications of
this result in section~\ref{sec:discussion}.

\section{\label{sec:weylccs} Weyl Consistency Conditions}
In general, the classical symmetries of a theory may be broken for its
renormalized Green functions.  The form of this ``anomaly'' is constrained
by the algebra of the symmetry group: for an infinitesimal transformation
generated by $\Delta^a$ acting on the   generating functional of
renormalized Green functions $\Gamma$, we have
\begin{equation}
\label{eq:wesszuminoccs}
\left[\Delta^a,\Delta^b \right] \Gamma = i f^{abc}
\Delta^c \Gamma,
\end{equation}
where  $f^{abc}$ are the structure constants of the symmetry group.  These
are the so-called Wess--Zumino consistency conditions~\cite{Wess:1971yu}.

It is useful to study a QFT on a curved background with spacetime-dependent
couplings so that the metric $\gamma_{\mu\nu} (x)$ and  couplings $g^I (x)$
act as sources for the stress-energy tensor and the operators (labelled by
$I$) in the Lagrangian, respectively. We only consider the case of
dimensionless couplings, so that in perturbation theory all the interaction
terms in the Lagrangian are nearly marginal. We introduce their
infinitesimal local Weyl transformations as
\begin{equation}
  \begin{aligned}
    \Delta_{\sigma} \gamma^{\mu\nu}(x) &=2\sigma (x)
      \gamma^{\mu\nu}(x)\,,\\
    \Delta_{\sigma} g^I(x) &=\sigma(x) \beta^I (x)\,,
  \end{aligned}
\end{equation}
where $\beta^I(x)$ is the beta function of the associated coupling and
depends on $x$ only through $g^I(x)$. The group of Weyl transformations is
Abelian and has only a single generator. Thus, Eq.~\eqref{eq:wesszuminoccs}
becomes
\begin{equation}
  \label{eq:weylccs}
  \left[ \Delta_{\sigma}, \Delta_{\sigma'} \right] \Gamma = 0,
\end{equation}
where it is understood that $\Gamma=\Gamma [\gamma^{\mu\nu}, g^I]$,
indicating the dependence on the metric and couplings as background fields.
If the flat-background theory is a CFT, \eqref{eq:weylccs} has been solved
in \cite{Deser:1993yx, Boulanger:2007st, Boulanger:2007ab}.

The response of $\Gamma$ to Weyl rescaling produces the Weyl anomaly
\begin{equation}
  \label{eq:weylanomaly}
  \begin{aligned}
    \Delta_{\sigma} \Gamma[\gamma^{\mu\nu}, g^I] &=
    \int d^{\hspace{1pt}d} x\sqrt{\gamma}\,\sigma
	\sum_i \left(a_i A_i[\gamma^{\mu\nu}]\right.\\
	&\left. \qquad\qquad+\, b_i B_i [\gamma^{\mu\nu}, g^I]
    + c_i C_i [g^I]\right),
  \end{aligned}
\end{equation}
where $d$ is the dimension of spacetime (presumed even here), and $i$ is a
counting index. The form of Eq.~\eqref{eq:weylanomaly} is fixed by general
diffeomorphism invariance and power counting.  $A_i, B_i$ and $C_i$ are
functions of the metric and couplings, and by dimensional analysis must
include $d$ spacetime derivatives.  The $A_i$ do not contain any
derivatives on couplings and are therefore of $d/2$-th order in curvature,
the $C_i$ are functions of $d$ derivatives on the couplings, and, finally,
the $B_i$ are functions of both †curvature and derivatives of the
couplings.  The coefficients $a_i$, $b_i$ and $c_i$ are all functions of
the couplings only. In particular, the $A_i$ contain the Euler term in $d$
dimensions with coefficient $(-1)^{d/2}a$, so that at fixed points $a>0$.

Now, the consistency conditions from Eq.~\eqref{eq:weylccs} impose
integrability relations on the terms in Eq.~\eqref{eq:weylanomaly}. The
relation of interest involves the coefficient of the Euler term in
Eq.~\eqref{eq:weylanomaly} and coefficients of terms in the $B_i$ involving
$H_{\mu\nu}$, a generalization of the Einstein tensor to $d$ dimensions
found by Lovelock~\cite{Lovelock:1971yv}. In even dimensions, it was shown
that an integrability relation exists~\cite{Grinstein:2013cka} involving
$a$ such that~\footnote{These $\chi_{IJ}$ and $w_I$ in $d=6$ were denoted
$\mathcal{H}_{IJ}^1$ and $\mathcal{H}_I^1$ respectively
in~\cite{Grinstein:2013cka}}
\begin{equation}
  \partial_I\tilde{a}=\frac{1}{d}(\chi_{IJ}+\partial_Iw_J
  -\partial_Jw_I)\beta^J,
  \label{eq:acc}
\end{equation}
which can be brought to the form
\begin{equation}
  \frac{d\tilde{a}}{d \log \mu}=\frac{1}{d}\chi_{IJ}\beta^I\beta^J,
  \label{eq:atheorem}
\end{equation}
where $\mu$ is the renormalization scale.  Here $\chi_{IJ}$ and $w_I$ are
tensors in the space of couplings and they appear in the coefficients of
the $B_i$ terms $\pd_{\mu} g^I \pd_{\nu} g^J H^{\mu\nu}$ and
$\nabla_\mu\partial_\nu g^I H^{\mu\nu}$ in Eq.~\eqref{eq:weylanomaly},
where $\tilde{a}$ is a scalar in the space of couplings~\footnote{For more
details the reader is referred to~\cite{Grinstein:2013cka}.}.  Both
quantities may be related to correlation functions of the stress-energy
tensor, its trace, and the operators in the QFT.  Since $\beta^I = 0$ at
the critical points, $\tilde{a}$ is stationary with respect to variations
of scale there. In fact
\begin{equation}
  \label{eq:aatilde}
  \tilde{a}=a+w_I\beta^I+\sum_p a_p,
\end{equation}
where $a_p$ are some of the $a_{i}$'s in \eqref{eq:weylanomaly} that vanish
at criticality. Hence, at critical points, $\tilde{a} = a$.  Moreover,
Eq.~\eqref{eq:atheorem} that $\tilde{a}$ satisfies is very similar to that
found for the analogous quantity in two dimensions
in~\cite{Zamolodchikov:1986gt}. This suggests $\tilde{a}$ as the analog of
Zamolodchikov's monotonically-decreasing function in two dimensions.

While the consistency conditions impose this integrability relation, a
strong version of the $a$-theorem must establish that the ``metric''
$\chi_{IJ}$ is positive-definite, which then proves that $d\tilde{a}/d\log
\mu > 0$. To compute $\chi_{IJ}$, other methods must be used.

\section{\label{sec:results}Results from the effective potential}
To compute $\chi_{IJ}$ in six dimensions, we work with the
conformally-coupled scalar field theory~\footnote{We do not study fermions
or vectors, which do not have interacting dynamics with classical scale
invariance at the perturbative level in six dimensions.} on a curved
background with Lagrangian
\begin{equation}
	\mathscr{L} = \tfrac{1}{2}(\pd_{\mu} \phi_i \pd_{\nu} \phi^i
	\gamma^{\mu\nu} + \tfrac{1}{5}R \phi_i \phi^i) + \tfrac{1}{3!}
	g_{ijk} \phi^i \phi^j \phi^k,
	\label{eq:6dlagrangian}
\end{equation}
with the fields, spacetime metric, and couplings all implicitly functions
of spacetime. The generic coupling constants $g^I$ are here specifically
$g_{ijk}$ with the label $I = (ijk)$.  At the classical level the term
$\pd_{\mu} g^I \pd_{\nu} g^J H^{\mu\nu}$, where the Lovelock tensor in
$d=6$ is
\begin{align*}
  H_{\mu\nu}&=(R^2 - 4R_{\kappa\lambda}R^{\kappa\lambda}+
  R_{\kappa\lambda\rho\sigma}R^{\kappa\lambda\rho\sigma})\gamma_{\mu\nu}\\
  &\quad-4 R R_{\mu\nu} + 8 R_{\mu\kappa}R^\kappa{\!}_\nu +
  8 R^{\kappa\lambda} R_{\kappa\mu\lambda\nu}\\
  &\quad-4 R_{\kappa\lambda\rho\mu}R^{\kappa\lambda\rho}{\!}_\nu,
\end{align*}
clearly does not show up, so $\chi_{IJ} = 0$ at the classical level. To
find the first (quantum) contributions to $\chi_{IJ}$, we can compute the
effective potential in a curved background with the loop expansion to two
loop order or, equivalently, second order in $\hbar$~\footnote{There is no
quantum correction to $\chi_{IJ}$ at one loop order. We work in units where
$\hbar = 1$ so that the loop-counting scheme is easier to use.}.

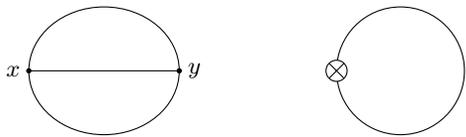
\begin{figure}[ht]
  \centering
  \begin{tikzpicture}
    \draw (0,0) circle [x radius=1cm, y radius=0.85cm];
    \draw (-1,0)--(1,0);
    \fill (-1,0) circle [radius=1pt] node[left] {$x$};
    \fill (1,0) circle [radius=1pt] node[right] {$y$};
  \end{tikzpicture}
  \hspace{1cm}
  \begin{tikzpicture}
    \draw (0,0) circle [radius=0.85cm];
    \filldraw[cross,fill=white] (-0.85,0) circle (4pt) node[left=2pt]
      {$x$};
  \end{tikzpicture}
  \caption{The diagrams that need to be considered at the two-loop level.}
  \label{fig:TwoLoopDiags}
\end{figure}

The six-dimensional two-loop effective potential can be computed using heat
kernel methods in dimensional regularization~\cite{Jack:1983sk,
Jack:1985wf, Kodaira:1985pg}.  This is done in position space, and it
involves the computation of the two-loop graph and the associated graph
with the counterterm insertion in Fig.~\ref{fig:TwoLoopDiags}. These two
graphs generate the full two-loop effective potential.
Such computations have been explained in great detail
in~\cite{Jack:1983sk}. The case of $d=6$ single-flavor $\phi^3$ theory with
$x$-independent coupling has been worked out in~\cite{Jack:1985wf,
Kodaira:1985pg}, and we find agreement with these papers in cases checked.
A detailed account of our computation, and further results not directly
pertinent to the conclusions of this Letter, will be presented in a
separate publication.

From our computation we determine the one- and two-loop
anomalous dimensions of the elementary fields $\phi_i$ and the beta
functions for the couplings $g_{ijk}$:
\begin{align}
\gamma^{(1)}&=\frac{1}{64\pi^3}\frac{1}{12}
\tikz[baseline=(vert_cent.base)]{
  \node (vert_cent) {\hspace{-13pt}$\phantom{-}$};
  \draw (0,0)--(0.3,0)
        (0.7,0) ++(0:0.4cm) arc (0:360:0.4cm and 0.3cm)
        (1.1,0)--(1.4,0);
},\\
\label{gammatwo}
\gamma^{(2)}&=\frac{1}{(64\pi^3)^2}\frac{1}{18}\left(
\tikz[baseline=(vert_cent.base)]{
  \node (vert_cent) {\hspace{-13pt}$\phantom{-}$};
  \draw (0,0)--(0.3,0)
        (0.7,0) ++(0:0.4cm) arc (0:360:0.4cm and 0.3cm)
        (0.7,0.3)--(0.7,-0.3)
        (1.1,0)--(1.4,0);
}
-\frac{11}{24}
\tikz[baseline=(vert_cent.base)]{
  \node (vert_cent) {\hspace{-13pt}$\phantom{-}$};
  \draw (0,0)--(0.3,0)
        (0.7,0) ++(0:0.4cm and 0.3cm) arc (0:50:0.4cm and 0.3cm) node (n1)
        {}
        (0.7,0) ++(50:0.4cm and 0.3cm) arc (50:130:0.4cm and 0.3cm) node
        (n2) {}
        (0.7,0) ++(130:0.4cm and 0.3cm) arc (130:360:0.4cm and 0.3cm)
        (n1.base) to[out=215,in=325] (n2.base)
        (1.1,0)--(1.4,0);
}
\right),\\
\label{betaone}
\beta^{(1)}&=-\frac{1}{64\pi^3}\left(
\tikz[baseline=(vert_cent.base)]{
  \node (vert_cent) {\hspace{-13pt}$\phantom{-}$};
  \draw (0.7,0) ++(0:0.35cm) arc (0:135:0.35cm) node (n1) {}
        (0.7,0) ++(135:0.35cm) arc (135:225:0.35cm) node (n2) {}
        (0.7,0) ++(225:0.35cm) arc (225:360:0.35cm) node (n3) {}
        (n1.base)--+(145:0.5cm)
        (n2.base)--+(215:0.5cm)
        (n3.base)--+(0:0.5cm);
}
-\frac{1}{12}
\tikz[baseline=(vert_cent.base)]{
  \node (vert_cent) {\hspace{-13pt}$\phantom{-}$};
  \draw (0,0.5)--(0.5,0)
        (0,-0.5)--(0.5,0)
        (0.5,0)--(0.8,0)
        (1.2,0) ++(0:0.4cm) arc (0:360:0.4cm and 0.3cm)
        (1.6,0)--(1.9,0);
}
\right),\\
\beta^{(2)}&=-\frac{1}{(64\pi^3)^2}\frac12\left(
\tikz[baseline=(vert_cent.base)]{
  \node (vert_cent) {\hspace{-13pt}$\phantom{-}$};
  \draw (0.7,0) ++(0:0.35cm) arc (0:45:0.35cm) node (n1) {}
        (0.7,0) ++(45:0.35cm) arc (45:135:0.35cm) node (n2) {};
  \draw[white] (0.7,0) ++(135:0.35cm) arc (135:225:0.35cm) node (n3) {};
  \draw (0.7,0) ++(225:0.35cm) arc (225:315:0.35cm) node (n4) {}
        (0.7,0) ++(315:0.35cm) arc (315:360:0.35cm) node (n5) {}
        (n2.base)--+(145:0.5cm)
        (n3.base)--+(215:0.5cm)
        (n5.base)--+(0:0.5cm);
  \draw[name path=a] (n1.base)--(n3.base);
  \draw[white, name path=b] (n2.base)--(n4.base);
  \path[name intersections={of=a and b,by=i}];
  \node[fill=white, inner sep=1pt, rotate=45] at (i) {};
  \draw (n2.base)--(n4.base);
}
-\frac{7}{36}
\tikz[baseline=(vert_cent.base)]{
  \node (vert_cent) {\hspace{-13pt}$\phantom{-}$};
  \draw (0.7,0) ++(0:0.35cm) arc (0:30:0.35cm) node (n1) {}
        (0.7,0) ++(30:0.35cm) arc (30:105:0.35cm) node (n2) {}
        (0.7,0) ++(105:0.35cm) arc (105:135:0.35cm) node (n3) {}
        (0.7,0) ++(135:0.35cm) arc (135:225:0.35cm) node (n4) {}
        (0.7,0) ++(225:0.35cm) arc (225:360:0.35cm) node (n5) {}
        (n1.base) to[out=200,in=300] (n2.base)
        (n3.base)--+(145:0.5cm)
        (n4.base)--+(215:0.5cm)
        (n5.base)--+(0:0.5cm);
}\right.\nonumber\\
\label{betatwo}
&\hspace{2cm}+\frac12
\tikz[baseline=(vert_cent.base)]{
  \node (vert_cent) {\hspace{-13pt}$\phantom{-}$};
  \draw (0.7,0) ++(0:0.35cm) arc (0:90:0.35cm) node (n1) {}
        (0.7,0) ++(90:0.35cm) arc (90:135:0.35cm) node (n2) {}
        (0.7,0) ++(135:0.35cm) arc (135:225:0.35cm) node (n3) {}
        (0.7,0) ++(225:0.35cm) arc (225:270:0.35cm) node (n4) {}
        (0.7,0) ++(270:0.35cm) arc (270:360:0.35cm) node (n5) {}
        (n1.base)--(n4.base)
        (n2.base)--+(145:0.5cm)
        (n3.base)--+(215:0.5cm)
        (n5.base)--+(0:0.5cm);
}
-\frac19
\tikz[baseline=(vert_cent.base)]{
  \node (vert_cent) {\hspace{-13pt}$\phantom{-}$};
  \draw (0,0.5)--(0.5,0)
        (0,-0.5)--(0.5,0)
        (0.5,0)--(0.8,0)
        (1.2,0) ++(0:0.4cm) arc (0:360:0.4cm and 0.3cm)
        (1.2,0.3)--(1.2,-0.3)
        (1.6,0)--(1.9,0);
}\\
&\nonumber\hspace{2cm}\left.+\frac{11}{216}
\tikz[baseline=(vert_cent.base)]{
  \node (vert_cent) {\hspace{-13pt}$\phantom{-}$};
  \draw (0,0.5)--(0.5,0)
        (0,-0.5)--(0.5,0)
        (0.5,0)--(0.8,0)
        (1.2,0) ++(0:0.4cm and 0.3cm) arc (0:50:0.4cm and 0.3cm) node (n1)
        {}
        (1.2,0) ++(50:0.4cm and 0.3cm) arc (50:130:0.4cm and 0.3cm) node
        (n2) {}
        (1.2,0) ++(130:0.4cm and 0.3cm) arc (130:360:0.4cm and 0.3cm)
        (n1.base) to[out=215,in=325] (n2.base)
        (1.6,0)--(1.9,0);
}
\right).
\end{align}
To our knowledge the multi-component two-loop results \eqref{gammatwo} and
\eqref{betatwo} have not appeared for general coupling $g_{ijk}$ before in
the literature, although they may be extracted from
Ref.~\cite{deAlcantaraBonfim:1981sy}. Here we have used  diagrammatic
notation to indicate the corresponding contraction of the couplings, {\it
e.g.},
\begin{equation}
\tikz[baseline=(vert_cent.base)]{
  \node (vert_cent) {\hspace{-13pt}$\phantom{-}$};
  \draw (0,0)--(0.3,0)
        (0.7,0) ++(0:0.4cm) arc (0:360:0.4cm and 0.3cm)
        (1.1,0)--(1.4,0);
}
=g_{ikl}g_{jkl},
\end{equation}
and permutations of the free indices in the wavefunction-renormalization
corrections to the beta function are understood. For example,
\begin{equation}
\tikz[baseline=(vert_cent.base)]{
  \node (vert_cent) {\hspace{-13pt}$\phantom{-}$};
  \draw (0,0.5)--(0.5,0)
        (0,-0.5)--(0.5,0)
        (0.5,0)--(0.8,0)
        (1.2,0) ++(0:0.4cm) arc (0:360:0.4cm and 0.3cm)
        (1.6,0)--(1.9,0);
}
=g_{ijl}g_{lmn}g_{kmn}+\text{permutations}.
\end{equation}
Eq.~\eqref{betaone} generalizes the single field result
of~\cite{Macfarlane:1974vp} (see also~\cite{Toms:1982af, Kodaira:1985vr,
Kodaira:1985pg, Jack:1985wf}) to the multi-field case, and agrees with the
results of \cite{Amit:1976pz, deAlcantaraBonfim:1980pe,
deAlcantaraBonfim:1981sy}. The first
contribution to \eqref{betatwo} is non-planar. For the seemingly asymmetric
vertex corrections in \eqref{betatwo} (the second and third terms) a
symmetrization is understood; for example,
\[
\tikz[baseline=(vert_cent.base)]{
  \node (vert_cent) {\hspace{-13pt}$\phantom{-}$};
  \draw (0.7,0) ++(0:0.35cm) arc (0:90:0.35cm) node (n1) {}
        (0.7,0) ++(90:0.35cm) arc (90:135:0.35cm) node (n2) {}
        (0.7,0) ++(135:0.35cm) arc (135:225:0.35cm) node (n3) {}
        (0.7,0) ++(225:0.35cm) arc (225:270:0.35cm) node (n4) {}
        (0.7,0) ++(270:0.35cm) arc (270:360:0.35cm) node (n5) {}
        (n1.base)--(n4.base)
        (n2.base)--+(145:0.5cm)
        (n3.base)--+(215:0.5cm)
        (n5.base)--+(0:0.5cm);
}
\quad\sim\quad
\tikz[baseline=(vert_cent.base)]{
  \node (vert_cent) {\hspace{-13pt}$\phantom{-}$};
  \draw (0.7,0) ++(0:0.35cm) arc (0:90:0.35cm) node (n1) {}
        (0.7,0) ++(90:0.35cm) arc (90:135:0.35cm) node (n2) {}
        (0.7,0) ++(135:0.35cm) arc (135:180:0.35cm) node (n3) {}
        (0.7,0) ++(180:0.35cm) arc (180:225:0.35cm) node (n4) {}
        (0.7,0) ++(225:0.35cm) arc (225:360:0.35cm) node (n5) {}
        (n1.base) to[out=270,in=0] (n3.base)
        (n2.base)--+(145:0.5cm)
        (n4.base)--+(215:0.5cm)
        (n5.base)--+(0:0.5cm);
}
+
\tikz[baseline=(vert_cent.base)]{
  \node (vert_cent) {\hspace{-13pt}$\phantom{-}$};
  \draw (0.7,0) ++(0:0.35cm) arc (0:90:0.35cm) node (n1) {}
        (0.7,0) ++(90:0.35cm) arc (90:135:0.35cm) node (n2) {}
        (0.7,0) ++(135:0.35cm) arc (135:225:0.35cm) node (n3) {}
        (0.7,0) ++(225:0.35cm) arc (225:270:0.35cm) node (n4) {}
        (0.7,0) ++(270:0.35cm) arc (270:360:0.35cm) node (n5) {}
        (n1.base)--(n4.base)
        (n2.base)--+(145:0.5cm)
        (n3.base)--+(215:0.5cm)
        (n5.base)--+(0:0.5cm);
}
+
\tikz[baseline=(vert_cent.base)]{
  \node (vert_cent) {\hspace{-13pt}$\phantom{-}$};
  \draw (0.7,0) ++(0:0.35cm) arc (0:135:0.35cm) node (n1) {}
        (0.7,0) ++(135:0.35cm) arc (135:180:0.35cm) node (n2) {}
        (0.7,0) ++(180:0.35cm) arc (180:225:0.35cm) node (n3) {}
        (0.7,0) ++(225:0.35cm) arc (225:270:0.35cm) node (n4) {}
        (0.7,0) ++(270:0.35cm) arc (270:360:0.35cm) node (n5) {}
        (n2.base) to[out=0,in=90] (n4.base)
        (n1.base)--+(145:0.5cm)
        (n3.base)--+(215:0.5cm)
        (n5.base)--+(0:0.5cm);
}\,
\]
where ``$\sim$'' means ``the left-hand side stands for the right-hand
side.''

Our main result is the two-loop expression for the ``metric'' in theory
space:
 \begin{equation}
   \chi_{IJ}^{(2)} = -\frac{1}{(64\pi^3)^2}\frac{1}{3240}\delta_{IJ}\,.
	\label{eq:twoloopchi}
\end{equation}
With this result and the one-loop beta function \eqref{betaone} we can use
the consistency condition \eqref{eq:acc} to compute $\tilde{a}$ at three
loops. For this we also need $w_I^{(2)}$, which we can obtain from the same
heat-kernel computation~\footnote{The fact that $w_I^{(2)}\sim g_I$
implies that $\partial_Iw_J^{(2)}-\partial_Jw_I^{(2)}=0$, and also, by
\eqref{eq:acc}, that the flow is gradient. In multicomponent systems in
four dimensions the gradient-property of the flow at low loop orders was
considered long ago in~\cite{Wallace:1974dy}}:
\begin{equation}
  w_I^{(2)}=-\frac{1}{(64\pi^3)^2}\frac{1}{12\hspace{1pt}960} g_I.
\end{equation}
We find~\footnote{With $n_\phi$ scalar fields there is also a contribution
to $\tilde{a}$ at zero coupling with $\tilde{a} =
\tfrac{1}{64\pi^3}n_\phi\tfrac{1}{9072}$ in the conventions
of~\cite{Grinstein:2013cka}, which we use here.}
\begin{equation}
\tilde{a}^{(3)} = \frac{1}{(64\pi^3)^3}\frac{1}{77\hspace{1pt}760}
\left(
\tikz[baseline=(vert_cent.base)]{
  \node (vert_cent) {\hspace{-13pt}$\phantom{-}$};
  \draw (0,0) circle [radius=0.35cm];
  \draw (0,0)--(120:0.35cm)
        (0,0)--(240:0.35cm)
        (0,0)--(0:0.35cm);
}-
\frac14\,
\tikz[baseline=(vert_cent.base)]{
  \node (vert_cent) {\hspace{-13pt}$\phantom{-}$};
  \draw (0,0) circle [x radius=0.45cm, y radius=0.35cm];
  \draw (-0.45,0)--(-0.2,0)
        (0.45,0)--(0.2,0);
  \draw (0,0) circle [x radius=0.2cm, y radius=0.15cm];
}
\right).
	\label{eq:threeloopa}
\end{equation}
The three-loop contribution to the coefficient of the Euler term $a$ can
also be computed using the relation between $\tilde{a}$ and $a$ of the form
\eqref{eq:aatilde} found in~\cite{Grinstein:2013cka}. We find
\begin{equation}
  a^{(3)}=\frac{1}{(64\pi^3)^3}\frac{1}{64\hspace{1pt}800}
  \left(
  \tikz[baseline=(vert_cent.base)]{
  \node (vert_cent) {\hspace{-13pt}$\phantom{-}$};
  \draw (0,0) circle [radius=0.35cm];
  \draw (0,0)--(120:0.35cm)
        (0,0)--(240:0.35cm)
        (0,0)--(0:0.35cm);
}-
\frac14\,
\tikz[baseline=(vert_cent.base)]{
  \node (vert_cent) {\hspace{-13pt}$\phantom{-}$};
  \draw (0,0) circle [x radius=0.45cm, y radius=0.35cm];
  \draw (-0.45,0)--(-0.2,0)
        (0.45,0)--(0.2,0);
  \draw (0,0) circle [x radius=0.2cm, y radius=0.15cm];
}
\right).
\end{equation}
Clearly, both $\tilde{a}$ and $a$ increase in the flow out of the trivial
fixed point.

One may wonder if the results in~\eqref{eq:twoloopchi}
and~\eqref{eq:threeloopa} depend on the renormalization scheme we used to
compute the two-loop effective potential. Actually, Eq.~\eqref{eq:acc} (and
thus Eq.~\eqref{eq:atheorem}) is invariant under the choice of
renormalization scheme.  The individual terms are, however,
scheme-dependent. The corresponding arbitrariness is of the form $\delta
\tilde a=z_{IJ}\beta^I\beta^J$ and $\delta \chi_{IJ}= \beta^K\partial_K
z_{IJ}+z_{KJ}\partial_I\beta^K +z_{IK}\partial_J\beta^K$, where $z_{IJ}$ is
an arbitrary regular symmetric function of the couplings. Since the
arbitrariness in $\tilde a$ vanishes (quadratically) when fixed points are
approached, it cannot change the nature of the flow in the vicinity of
fixed points.

\section{\label{sec:discussion}Discussion}
Using the result of our computation, Eq.~\eqref{eq:twoloopchi}, in the
evolution equation \eqref{eq:atheorem}, or equivalently, the explicit form
of $\tilde a$ in \eqref{eq:threeloopa}, it is apparent that in perturbation
theory the quantity $\tilde{a}$ in Eq.~\eqref{eq:atheorem} actually {\em
increases} as one decreases the renormalization scale.  This is contrary to
intuition developed in $d = 2, 4$, where $\tilde{a}$ seems to count the
degrees of freedom in a QFT.

This result should be taken with two comments in mind. Firstly, that the
result is a perturbative one, and we cannot say anything about
non-perturbative regimes of six-dimensional QFTs. And secondly, that there
are no known perturbative critical points other than the single, trivial
one at $g_{ijk} = 0$, so in this context renormalization group flows do not
connect pairs of critical points~\footnote{This does not mean that they do
not exist. Non-trivial, perturbative flows between a UV and IR critical
point have been studied in $6-\epsilon$ dimensions in the $O(N)$ model
recently, as in~\cite{Fei:2014yja}. It is an open question as to whether or
not such results could be extended to six dimensions.}. However, it is
still true that, with Eq.~\eqref{eq:atheorem} identical in $d = 2, 4$, and
$6$ dimensions, the strong version of the $a$-theorem holds perturbatively
in $d = 2, 4$ but not in $d = 6$ \footnote{Note that, as suggested
in~\cite{Elvang:2012st}, the weak version of the $a$-theorem may still be
true in $d=6$ for flows at weak coupling connecting two critical points.}.

We do not know the reason for this difference. One possibility may be the
unstable nature of the theory we are considering. After all, a cubic
potential is unbounded from below. However, the state with $\langle
\phi_i(x)\rangle = 0$ is perturbatively stable and our computations are
valid only in the perturbative regime. Moreover, the analogous case in four
dimensions, the inverted quartic potential, is also unstable, but does
satisfy a perturbative $a$-theorem (since the metric in theory space,
$\chi_{IJ}$, is perturbatively positive in four dimensions, independently
of the sign of the quartic couplings).  Another possibility is that a flow
between critical points is required for an $a$-theorem to hold, but the
only perturbatively-accessible critical point in the class of theories in
Eq.~\eqref{eq:6dlagrangian} is the Gaussian fixed point at $g_{ijk}=0$.
But, again comparing to known cases, a perturbative strong $a$-theorem
holds for scalar theories in four dimensions, in spite of only having a
Gaussian fixed point at the origin of coupling-constant space.

$a$-theorems can be used to restrict proposed dynamics of strongly
interacting models~\cite{Cardy:1988cwa}. If our result that $\tilde a$
increases in flows towards the IR holds even non-perturbatively, one could
envision using it to restrict putative dynamics of strongly interacting
QFTs in $d=6$.  In this sense, the existence of an ``anti-$a$-theorem'' may
be just as useful as a normal one. It is therefore of interest to
investigate renormalization group flows in the vicinity of non-Lagrangian
critical QFTs that have been formulated through studies of low energy
dynamics of M5 branes. Of course, another avenue of research is the
establishment of the theorem non-perturbatively in the presence of a flow
between fixed points.

Finally, let us note that there may be quantities that reduce to $a$ at
fixed points that are not of the form of $\tilde{a}$ (up to the ambiguity
$z_{IJ}\beta^I\beta^J$), but that do undergo monotonically-decreasing RG
flow towards the IR. This possibility was explored
in~\cite{Yonekura:2012kb}.
\vspace{20pt}

\begin{acknowledgements}
  For our computations we relied heavily on \emph{Mathematica} and the
  package \href{http://www.xact.es/}{\texttt{xAct}}. We would like to thank
  Tom Appelquist, George Fleming, Ken Intriligator, Hugh Osborn, David
  Poland, and Jaewon Song for valuable discussions. We also thank Ken
  Intriligator and Hugh Osborn for comments on the manuscript. Finally, we
  thank J.~A.~Gracey and one of the referees for bringing references
  \cite{deAlcantaraBonfim:1980pe, deAlcantaraBonfim:1981sy} to our
  attention.

  The work of BG, DS and MZ was supported in part by the US Department of
  Energy under contract DE-SC0009919. AS thanks the KITP for its
  hospitality during the completion of this work. The research of AS is
  supported in part by the National Science Foundation under Grant No.\
  1350180. MZ would like to thank the support from the National Natural
  Science Foundation of China under grants No.\ 11105223 and No.\ 11205242.
\end{acknowledgements}

\bibliography{PRL_SixD}

\end{document}